\documentclass[11pt]{article}

\usepackage{graphicx}  
\usepackage{dcolumn}   
\usepackage{bm}        
\usepackage[english]{babel}
\usepackage{amsfonts,amsmath,amssymb,mathrsfs}
\usepackage{subfigure}
\usepackage{color}
\usepackage{hyperref}

\usepackage[left=2.5cm,top=3cm]{geometry} 

\usepackage{ulem} 

\def\a{\alpha}
\def\r{\rho}
\def\s{\sigma}
\def\t{\tau}
\def\m{\mu}
\def\n{\nu}
\def\k{\kappa}
\def\th{\theta}
\def\g{\gamma}\def\G{\Gamma}
\def\L{\Lambda}\def\l{\lambda}
\def\del{\nabla}
\def\D{\Delta}
\def\la{\langle}
\def\ra{\rangle}
\def\o{\omega}\def\O{\Omega}
\def\d{\delta}
\def\p{\partial}

\def \rh {\sqrt {-h} \,}

\def\half{\textstyle{\frac{1}{2}}}

\def\bdoc{\begin{document}}
\def\edoc{\end{document}}

\def\beq{\begin{equation}}
\def\eeq{\end{equation}}
\def\bea{\begin{eqnarray}}
\def\eea{\end{eqnarray}}
\def\ben{\begin{enumerate}}
\def\een{\end{enumerate}}
\def\la{\langle}
\def\ra{\rangle}
\def\a{\alpha}
\def\b{\beta}
\def\g{\gamma}\def\G{\Gamma}
\def\d{\delta}\def\D{\Delta}
\def\e{\epsilon}
\def\z{\zeta}
\def \f {\frac}
\def\th{\theta}
\def\k{\kappa}
\def\l{\lambda}
\def\m{\mu}
\def\n{\nu}
\def\o{\omega}
\def\p{\pi}
\def\r{\rho}
\def\s{\sigma}
\def\t{\tau}
\def\L{${\cal L}$}
\def\H{${\cal H}$}
\def\S{\Sigma }
\def\gsim{\; \raisebox{-.8ex}{$\stackrel{\textstyle >}{\sim}$}\;}
\def\lsim{\; \raisebox{-.8ex}{$\stackrel{\textstyle <}{\sim}$}\;}
\def\gtrsim{\gsim}
\def\lessim{\lsim}
\def\loc{{\rm local}}
\def\vm{v_{\rm max}}
\def\bh{\bar{h}}
\def\del{\partial}
\def\nab{\nabla}
\def\half{{\textstyle{\frac{1}{2}}}}
\def\fourth{{\textstyle{\frac{1}{4}}}}

\def\bD{{\bf D}}
\def\bE{{\bf E}}
\def\bF{{\bf F}}
\def\bB{{\bf B}}
\def\bP{{\bf P}}
\def\bV{{\bf v}}
\def\bv{{\bf v}}
\def\bx{{\bf x}}
\def\by{{\bf y}}
\def\bz{{\bf z}}
\def\ba{{\bf a}}
\def\bd{{\bf d}}
\def\bs{{\bf s}}
\def\bn{{\bf n}}
\def\bp{{\bf p}}

\def\O{\Omega}

\def\del{\nabla}
\def\br{{\bf r}}
\def\bnab{{\bf \nab}}
\def\lf{\left (}
\def\rt{\right)}
\def\tE{\tilde{E}}
\def\tL{\tilde{L}}
\def\Horava{Ho\v{r}ava }
\def\le{\left }
\def\re{\right}

\begin{document}

\title{The Black Hole Membrane Paradigm in $f(R)$ Gravity}
\author{Saugata Chatterjee\footnote{saugata@iucaa.ernet.in}\vspace*{0.1in} \\
{\footnotesize \it Inter-University Centre for Astronomy and Astrophysics (IUCAA)} \\
{\footnotesize \it Post Bag 4, Ganeshkhind, Pune 411~007, India} \vspace*{0.1in} \\ 
Maulik Parikh\footnote{maulik.parikh@asu.edu}\vspace*{0.1in} \\
{\footnotesize \it Department of Physics and Beyond: Center for Fundamental Concepts in Science}\\{\footnotesize \it Arizona State University, Tempe, AZ 85287, USA}\vspace*{0.1in} \\
Sudipta Sarkar\footnote{sudipta@umd.edu}\vspace*{0.1in} \\ 
{\footnotesize \it Department of Physics, University of Maryland}\\{\footnotesize \it College Park, MD 20742, USA}\vspace*{0.1in} \\ 
}
\date{\today} 
\maketitle
\thispagestyle{empty}

\begin{abstract}
\noindent
To an outside observer, a black hole's event horizon appears to behave exactly like a dynamical fluid membrane. We extend this membrane paradigm to black holes in general $f(R)$ theories of gravity. We derive the stress tensor and various transport coefficients of the fluid and find that the membrane behaves as a non-Newtonian fluid except for the special case of Einstein gravity. Using Euclidean methods, we study the thermodynamics of the membrane. We speculate on what theories of gravity admit horizons with fluid properties.
\end{abstract}

\newpage
\section{Introduction}

The membrane paradigm is the surprising idea that, to an outside observer, black hole horizons behave like fluid membranes. That is, when a black hole is perturbed by external fields, the equations of motion describing the response of the horizon are exactly what they would be if the fields were interacting instead with a bubble, or membrane, enveloping the horizon. The membrane is endowed with the sources for whatever external fields are present. In particular, to source the gravitational field, the membrane possesses the stress tensor of a viscous fluid. This external perspective of horizons as fluid membranes provides not only an intuitive way of understanding black hole interactions but also the original semiclassical realization of holography.

The membrane paradigm was first discovered \cite{thorne,damour} by re-writing particular field equations of perturbed black hole horizons in terms of familiar nonrelativistic dissipative equations such as Ohm's law and the Navier-Stokes equation. A more systematic action-based derivation was obtained in \cite{membrane}, which in principle enabled membrane properties to be determined for arbitrary field theories. Nevertheless, many puzzles remain. For what gravitational theories does a black hole horizon behave as if it were a Newtonian fluid? What are the fluid transport coefficients in more general theories of gravity? Does the membrane always obey the Navier-Stokes equation? And more generally, for what gravitational theories does the membrane paradigm even exist? In this paper, we attempt to shed some light on these questions by considering the action formulation of the membrane paradigm for $f(R)$ theories of gravity. These theories serve as a model for higher-derivative gravity; they are a simple extension of Einstein gravity in that they introduce exactly one extra degree of freedom.
\section{Geometric set-up}

Before entering into the details of the membrane action, it will help to specify precisely the geometric set-up.
We will work in $D$ spacetime dimensions. Let the black hole event horizon, $H$, which is a $D\!-\!1$-dimensional null hypersurface, be generated by null geodesics $l^a$. We take these generators to have a nonaffine parameterization, $\tau$. That is, $l^a = (\partial/\partial \tau)^a$ and the geodesic equation is $l^a \nabla_a l^b = \kappa l^b$, rather than zero. Here $\kappa$ is a nonaffine coefficient; for a stationary spacetime, $l^a$ coincides with the null limit of the timelike Killing vector and $\kappa$ can then be interpreted as the surface gravity of the horizon.

Although the membrane paradigm can be formulated entirely on the event horizon, it proves convenient to introduce a timelike stretched horizon, $\Sigma$, positioned slightly outside $H$, the advantage being that a timelike surface has a nondegenerate metric which permits one to write down a conventional action. The precise choice of the timelike surface is somewhat arbitrary. We consider $\Sigma$ to be one among a foliation of timelike surfaces, each labeled by a parameter $\alpha$ such that in the limit $\alpha \to 0$, the stretched horizon approaches the true horizon. In the absence of horizon caustics, a one-to-one correspondence between points on $H$ and $\Sigma$ are always possible by, for example, using ingoing light rays that connect both the surfaces.


We can also regard the stretched horizon as the world-tube of a family of ``fiducial" observers hovering just outside the black hole. We take these observers to have world lines $U^a$;
then just as $H$ is generated by the null congruence $l^a$, the stretched horizon is generated by the timelike congruence $U^a$. The stretched horizon also has a spacelike, outward pointing unit normal vector $n^a$. 

In the limit $ \alpha \to 0$, we require that $\alpha U^a \to l^a$ and $\alpha n^a \to l^a$ i.e. the stretched horizon tends to the true horizon in this limit, as we have already envisaged. This is nothing more than the statement that the null generator $l^a$ is both normal and tangential to the true horizon, which is the defining property of null surfaces. The metric $h_{ab}$ on the stretched horizon $\Sigma$ can be expressed in terms of the spacetime metric $g_{ab}$ and the normal vector $n^a$. Similarly we can define a metric $\gamma_{ab}$ on a $D-2$-dimensional spacelike cross-section of $\Sigma$, to which $U^a$ is normal:
\bea
h_{ab} = g_{ab} - n_a n_b ~~ \textrm{and} ~~ \gamma_{ab} = h_{ab} + U_a U_b.
\eea
We will choose the stretched horizon among all possible choices, such that the normal vector $n^a$ obeys an affine geodesic equation, $n^a \nabla_a n_b = 0$,
and as a result, for any vector $v^a \in \Sigma$, we have $\nabla_a v^a = v^{a}_{|a}$ where $_{|a}$ is the covariant derivative with respect to the metric $h_{ab}$. Next, we denote $\{A,B,...\}$ as the coordinates on the $D\!-\!2$-dimensional spacelike cross-section of \H ~and $k^{A}_{B} = \gamma^{d}_{B} l^{A}_{|| d}$ as the extrinsic curvature on the $D\!-\!2$-dimensional cross-section of the null surface, where $_{||A}$ is the covariant derivative with respect to $\gamma_{AB}$.

We define the extrinsic curvature of $\Sigma$ as $ K^{a}_{b} = h^{c}_{b}\nabla_c n^a$. In the null limit $\alpha \to 0$, the various components of the extrinsic curvature become \cite{thorne}
\bea
\textrm{As $\alpha \to 0$}&:& K^{U}_{U} = K^{a}_{b} U_a U^b  \to - \alpha^{-1} \kappa \nonumber \\
&& K^{U}_{A} \to 0; K^{A}_{B} \to \alpha^{-1} k^{A}_{B}\nonumber \\
&& K \to \alpha^{-1} (\theta + \kappa) \label{nulllimit}
\eea
where $\theta$ is the expansion scalar of the geodesic congruence generating the horizon. Note that, for a $D\!-\!1$-dimensional timelike hypersurface, the extrinsic curvature of a $D\!-\!2$-dimensional spacelike section with respect to its timelike normal $U^a$ within the hypersurface has nothing to do with the (projection of the) extrinsic curvature $K^A_B$ with respect to the spacelike normal $n^a$ off the hypersurface. However, in the null limit, both $U^a$ and $n^a$ map to the same null vector $l^a$. Hence we have $K^A_B \to \alpha^{-1} k^A_B$ where $k^A_B$ is the the extrinsic curvature of a $D\!-\!2$-dimensional spacelike section of the horizon. We can then decompose $k_{AB}$ into a traceless part and a trace as
\begin{eqnarray}
k_{AB} = \sigma_{AB} +  \frac{1}{D-2} \theta \gamma_{AB} \nonumber 
\end{eqnarray}
where $\sigma_{AB}$ is the shear of the null congruence. In the null limit, various components of the extrinsic curvature diverge and we need to renormalize them by multiplying by a factor of $\alpha$. The physical reason behind such infinities is that, as the stretched horizon approaches the true one, the fiducial observers experience more and more gravitational blue shift; on the true horizon, the amount of blue shift is infinite. 
This completes the description of our geometric set-up. Next, we review the derivation of the black hole membrane paradigm in standard Einstein gravity.

\section{The membrane paradigm in Einstein gravity}
Since the region inside the event horizon cannot classically affect an outside observer, the classical equations of motion for such an observer must follow from the variation of the action restricted to the spacetime external to the black hole. 
However, the external action, $S^{\rm out}$, is not stationary on its own because boundary conditions are not fixed at the horizon, and hence the boundary term in the derivation of the Euler-Lagrange equations does not vanish at the horizon as it does at infinity. In order to obtain the correct equations, we must add a surface term to the action whose variation cancels this residual boundary variation. We do this by splitting the action as
\bea
S = \lf S^{\rm{in}} - S^{\rm{surf}} \rt + \lf S^{\rm{out}} + S^{\rm{surf}} \rt \; ,
\eea   
where $S^{\rm{surf}}$ is the requisite surface term, chosen so that $\delta S^{\textrm{out}} + \delta S^{\rm{surf}} = 0$. The surface term corresponds to sources such as surface electric charges and currents for the Maxwell action, or a surface energy-momentum tensor for the gravitational action. These sources are fictitious in a traditional sense because an infalling observer passing through the horizon would not detect them. Nevertheless, to the external observer they are very much real and observable. An ontologically different stance, as advocated by the principle of observer complementarity, is that both the infalling and external viewpoints are equally valid, even though they seemingly contradict each other; indeed, the infalling observer is unable to detect Hawking radiation either but that is not usually regarded as implying that Hawking radiation is fictitious.

For Einstein gravity, the external action is given by
\bea
S^{\rm{out}} = \frac{1}{16 \pi G} \int d^{D} x \sqrt{-g}~ R + \frac{1}{8\pi G} \oint_{\infty} d^{D-1} x \sqrt{h}~ K \; ,
\eea
where the second term is the Gibbons-Hawking boundary term required to cancel the normal derivatives of the variation of metric on the boundary at infinity. 
As before, extremizing this action does not yield the Einstein equations because of variational contributions at the (stretched) horizon.
To cancel this contribution, we add a surface term $S^{\rm{surf}}$ whose variation,
\bea
\delta S^{\rm{surf}} = \frac{1}{2} \int d^{D-1}x \sqrt{h} \, t^{ab} \, \delta h_{ab} \; ,
\eea
defines a surface energy-momentum tensor on the stretched horizon. This can be shown \cite{membrane} to take the form
\bea
t^{ab} = \frac{1}{8\pi G} \lf K h^{ab} - K^{ab}  \rt \label{stresstensorGR} \; ,
\eea
where $K_{ab}$ is the extrinsic curvature of the stretched horizon.
By invoking the Gauss-Codazzi equations, the energy-momentum tensor can be shown to satisfy a conservation equation:
\bea
t^{ab}_{|b} = - h^{a}_{c} T^{cd} n_{d} \; , \label{conserve}
\eea
where $T^{ab}$ is the energy-momentum tensor of real (bulk) matter outside the black hole. This is a continuity equation; it indicates that the divergence of the horizon energy-momentum tensor is equal to the flow of outside matter on to the horizon. The fact that the horizon energy-momentum tensor participates in a continuity equation with actual outside matter is crucial in sustaining the outside observer's belief that the surface energy-momentum tensor describes real matter.

In the limit that the stretched horizon approaches the true horizon, we can use \ref{nulllimit} to express the regularized stress tensor in terms of the horizon expansion and shear. Remarkably, the stress tensor projected on a $D\!-\!2$-dimensional cross-section of the horizon then takes the form of the stress tensor of a viscous fluid \cite{thorne,membrane,eling}:
\bea
t^{A}_{B} = p \gamma^{A}_{B} - 2 \eta \sigma^{A}_{B} - \zeta \theta \gamma^{A}_{B} \; ,
\eea
where $p = \frac{\kappa}{8\pi G}$ is the pressure, $\eta = \frac{1}{16 \pi G}$ the shear viscosity, and $\zeta = -  \frac{1}{8\pi G} \frac{D-3}{D-2}$ the bulk viscosity of the membrane. The constancy of the transport coefficients means that the event horizon behaves as a $D\!-\!2$-dimensional Newtonian fluid. Note that,
unlike ordinary fluids, the membrane has negative bulk viscosity. This would ordinarily
indicate an instability against generic perturbations triggering expansion or contraction. It
can be regarded as reflecting a null hypersurface's natural tendency to expand or contract.

Inserting the energy density $t^{a}_{b}U_a U^b = \Sigma = \alpha^{-1} \Sigma_{{\cal H}}$ into the conservation equation of the membrane stress tensor, we find that
\bea
\frac{d \Sigma_{{\cal H}}}{d \tau} + \Sigma_{{\cal H}} \theta = - p\theta + \zeta \theta^2 + 2 \eta \sigma^2 + T^{a}_{b} l_a l^b \label{energyconservGR} \; .
\eea
This is again the same as the energy conservation equation of a fluid with pressure $p$, shear viscosity $\eta$, and bulk viscosity $\zeta$ \cite{Landau}. Next, inserting the $A$th-momentum density, $\pi_A = t^{b}_{a} \gamma^{a}_{A}U_{b}$, into the conservation equation of the membrane stress tensor, we arrive at the momentum conservation equation of the membrane:
\bea
{\cal L}_{l^a}~ \pi_A + \theta~\pi_{A} &=& -  \nabla_{A} p  + 2 \eta \sigma^{B}_{A ||B } + \zeta \nabla_{A} \theta + T_{A}^{l} \label{NVforGR} \; ,
\eea
where ${\cal L}_{l^a}$ is the Lie derivative along the null direction. Since the Lie derivative along a congruence plays the role of the convective derivative in ordinary fluid dynamics, we recognize this as the Navier-Stokes equation of a viscous fluid. This completes our short review of the membrane paradigm in Einstein gravity. Next, we turn to its extension to $f(R)$ gravity.

\section{Extension to $f(R)$ gravity }
Consider a general diffeomorphism-invariant action of the form
\bea
 S= \frac{1}{16 \pi G} \int d^{D} x~ L (g_{ab}, R_{abcd}) + S_{\infty} + S_{\rm matter} \; ,
 \eea
where $S_\infty$ is the appropriate generalization of the Gibbons-Hawking term at infinity, whose precise form we do not need. 
After variation, the surface term on the stretched horizon is
\bea
\delta S_{\Sigma} &=&  { 1 \over 8 \pi G } \int d^{D-1}x \sqrt{-h} \, n_a \left[ P^{abcd} \nabla_d \delta g_{bc} - \delta g_{bc} \nabla_d P^{abcd} \right]   \; ,
\eea
where the tensor $P^{abcd} = \partial L/\partial R_{abcd}$ has all the symmetry properties of the Riemann curvature tensor. To proceed further, we specialize to the particular case for which the Lagrangian is a function $f(R)$ of the Ricci scalar only. The equation of motion for $f(R)$ gravity is
\bea
f'(R) R_{ab} - \del_a \del_b f'(R) + \left(\Box f'(R) - \frac{1}{2} f'(R) \right) g_{ab} = 8 \pi G ~T_{ab}\label{eom} \; ,
\eea
where the prime denotes a derivative with respect to the argument; when $f(R) = R$, this reduces to Einstein's equation. In order to obtain this equation, we need to add a surface term to the action, with variation
\bea
\delta S^{\rm surf}_{\Sigma} &=&-  { 1 \over 8 \pi G } \int d^{D-1}x \sqrt{-h} \left(F_1 + F_2\right)\label{variationaction_g} \; ,
\eea 
where, using $P^{abcd} = \frac{1}{2}( g^{ac} g^{bd} - g^{ad} g^{bc}) f'(R)$,
the terms $F_1$ and $F_2$ are given by,\footnote{We have used the gauge choice $ \delta n_a =0 $, which implies $ g^{ab} \delta g_{bc} = h^{ab} \delta g_{bc} = g^{ab} \delta h_{bc} =  h^{ab} \delta h_{bc} $.}
\bea
F_1 &=& \frac{1}{2}  h^{bc} \left[ \nabla_c \left( f'(R) n^a \delta g_{ab} \right) - \nabla_a   \left( f'(R) n^a \delta g_{bc} \right)  \right] \\
F_2 &=& - \frac{1}{2} h^{bd} \left[ \nabla_d n^c f'(R) + 2 n^c \nabla_d f'(R)  \right]\delta g_{bc} + h^{bc} \left[ f'(R) \nabla_a n^a + 2 n^a \nabla_a f'(R) \right]\delta g_{bc}
\eea
The contribution from the term $F_1$ can be shown to vanish in the limit in which the stretched horizon approaches the event horizon.
After some straightforward manipulation we find that
\bea
\delta S^{\rm surf}_{\Sigma} &=& -{ 1 \over 16 \pi G } \int d^{D-1}x \, \sqrt{-h} \, \left[ f'(R) \lf K h^{ab} - K^{ab}\rt + 2 n^d \del_d f'(R) h^{ab} \right] \, \delta h_{ab} \nonumber \\
& \equiv & -{ 1 \over 2 } \int d^{D-1}x \, \sqrt{-h} \, t^{ab} \, \delta h_{ab} 
\eea 
where $t^{ab}$ is the membrane stress tensor:
\bea
t^{ab} = \frac{1}{8\pi G} \left[ f'(R) \lf K h^{ab} - K^{ab}\rt + 2 n^d \del_d f'(R) h^{ab} \right] \label{stresstensor} \; .
\eea

This is the stress tensor for the membrane in $f(R)$ gravity. However, taking its divergence does not give a conservation equation analogous to (\ref{conserve}). This would seem to undermine the interpretation of $t^{ab}$ as real energy-momentum, which an observer would naturally require to be conserved. One way out is to note that the membrane stress tensor produces a discontinuity in the extrinsic curvature across the stretched horizon. The relationship between the discontinuity and the source term is given by the appropriate Israel junction condition \cite{Deruelle} for $f(R)$ gravity, which is
\bea
| f'(R) \lf K h_{ab} - K_{ab} \rt + h_{ab} n^d \del_d f'(R) |= 8 \pi G~  t_{ab}\label{junctioncond} \; ,
\eea
where $|A| \equiv  A_{+} - A_{-}$ denotes the difference between the quantities evaluated on the stretched horizon between its embedding in the external universe and in the spacetime internal to the black hole.  Comparing the junction condition (\ref{junctioncond}) with the membrane stress tensor (\ref{stresstensor}), we find
\bea
f'(R) \left( K h_{ab} - K_{ab} \right) + h_{ab} n^d \del_d f'(R)|_{-} +   h_{ab} n^d \del_d f'(R)|_{+} = 0 \label{condition} \; .
\eea
Now, junction conditions for a general $f(R)$ gravity theory have to be supplemented with an additional condition namely, the continuity of the Ricci scalar across the junction \cite{Deruelle}.
The reason behind this extra constraint is that, unlike for Einstein gravity, the equation of motion of a general $f(R)$ gravity is of fourth order. Unless the continuity of the Ricci scalar is imposed on the junction, the junction conditions do not reduce to the familiar Israel junction conditions as $f(R) \to R$. 
Another alternative but equivalent way to understand this extra constraint is that any $f(R)$ theory, other than Einstein gravity, can be cast into a scalar-tensor theory via a conformal transformation. Thus, apart from the tensor mode, a general $f(R)$ gravity also contains an extra scalar degree of freedom; it is the Ricci scalar that plays the role of this scalar field in the scalar-tensor picture. Hence, the continuity of the Ricci scalar actually ensures that the scalar degree of freedom is continuous across the junction. On the other hand, 
it is not possible to write down a conserved membrane source for scalar field theory due to the absence of a conserved current. By imposing this condition on our membrane, we are thereby effectively removing the scalar degree of freedom. The continuity of $R$ across the membrane leads to the continuity of the trace of extrinsic curvature $K$ \cite{Deruelle}.  Using this, (\ref{eom}) and (\ref{condition}), we find that
\bea
t^{ab}_{|b} = - h^{a}_{c} T^{cd} n_d \label{conservation} \; .
\eea
This is once again a conservation law.

Although the use of the junction conditions unambiguously leads to the correct conservation law, there is still something dissatisfying about it. The whole idea of the membrane paradigm is that we should not have to consider conditions on the other side of the membrane; using junction conditions does not seem to fit that philosophy. It would be nice to find another motivation for (\ref{conservation}). One intriguing possibility arises from observing that, for Einstein gravity, (\ref{stresstensorGR}) is simply the momentum of gravity in a Hamiltonian picture where ``time" runs in a spacelike direction off the stretched horizon. The existence of the conservation equation (\ref{conserve}) can then be recast as the momentum constraint equation. That in turn arises because of gauge invariance. This viewpoint explains why scalar field theories do not have a realistic intepretation in terms of the membrane paradigm. Applied to $f(R)$ theory it suggests that, since the theory does possess diffeomorphism invariance, there must exist some kind of conservation equation for the membrane stress tensor. It could be very illuminating to make these ideas more precise. In particular, it suggests that Lovelock theories, which have the same number of degrees of freedom as Einstein gravity, might admit a very clean interpretation in terms of fluid membranes.

The projection of the membrane stress tensor, (\ref{stresstensor}), to a spatial $(D\!-\!2)$-dimensional slice of the horizon gives
\bea
t^{AB} = \frac{1}{8\pi G} \left[ \lf \kappa f'(R) + 2 \frac{d f'(R)}{d \tau} \rt \gamma^{AB} + \frac{(D-3)}{(D-2)}\theta f'(R) \gamma^{AB} - f'(R) \sigma^{AB} \right]
\eea
where $\tau$ is the nonaffine parameter of the null congruence and $l^a \del_a = d  / d\tau$. The stress tensor resembles that of a viscous fluid and readily allows us to find the fluid transport coefficients:
\bea
\textrm{Pressure}:  p &=&  \frac{1}{8\pi G}\lf \kappa f'(R) + 2 \frac{d f'(R)}{d \tau} \rt \nonumber \\
\textrm{Shear viscosity}: \eta &=& \frac{f'(R)}{16\pi G} \nonumber \\
\textrm{Bulk Viscosity}:  \zeta &=& - \frac{(D-3)}{(D-2)}\frac{f'(R)}{8 \pi G} \nonumber \\
\textrm{Energy Density}:  \Sigma_{{\cal H}} &=&  \frac{1}{8\pi G} \left[ - \theta f'(R) - 2 \frac{d f'(R)}{d \tau} \right]
\eea
As in Einstein gravity, then, the membrane stress tensor for any general $f(R)$ gravity can indeed be written as a fluid stress tensor. However, there are a few differences:
\begin{itemize}

\item Transport coefficients are not constants but depend on the flow, characteristic of a non-Newtonian fluid \cite{fluidbooks}. 


\item For any $f(R)$ theory, the bulk viscosity coefficient is always negative provided $D > 3$. As in Einstein gravity, this is related to the teleological definition of the event horizon; it is independent of the theory of gravity.

\item Since, the transport coefficients are not constants, the relevant Navier-Stokes equation, which involves the derivative of the stress tensor, will have extra terms proportional to the derivatives of these transport coefficients. For fluid with constant viscosity, such terms do not contribute and we obtain the regular Navier-Stokes equation\footnote{The derivation of the Navier-Stokes equation usually assumes the constancy of various transport coefficients. But it is straightforward to lift that assumption and derive a general Navier-Stokes equation for a fluid with nonconstant viscosity; see, e.g., section $15$ in \cite{Landau}}. 

\end{itemize}
Another important property of the membrane fluid is the saturation of the so-called KSS bound \cite{kss}. As in Einstein gravity, the ratio of the shear viscosity to entropy per unit area is $1/4\pi$. In fact, in the context of finite temperature AdS/CFT, it was suggested that, for any $f(R)$ gravity theory, the ratio $\eta / s$ always saturates the KSS bound \cite{brustein}. Our result confirms this for the membrane fluid as well.

We can cast the conservation law into equations of fluid dynamics. Inserting the A-momentum density $\pi_A = t^{b}_{a} \gamma^{a}_{A}U_{b}$ into the conservation equation of the membrane stress tensor, we arrive at the momentum conservation equation of the membrane as:
\bea
{\cal L}_{l} \pi_A + \theta \pi_{A} &=& -  \nabla_{A} p  + 2 \lf \eta \sigma^B_A\rt _{||B} + \zeta \nabla_{A} \theta + T_{A}^{l} + \theta  \gamma_{A}^{B} \nabla_B \zeta 
\eea
This is identical to the Navier-Stokes equation of a viscous fluid provided we generalize the usual Navier-Stokes equation for the case of non-Newtonian fluids with nonconstant transport coefficients \cite{Landau}.  As a result, compared to Eq. (\ref{NVforGR}), this equation has extra terms involving the change of transport coefficients along the flow. In Einstein gravity, all such terms vanish since the viscous transport coefficients in that case are constants. 

Let us summarize the broad picture as follows: we have derived the membrane stress tensor for a general $f(R)$ gravity theory and proved that the stress tensor behaves like that of a non-Newtonian viscous fluid provided we imposed the continuity of the scalar curvature across the membrane. This condition can be justified both by imposing the junction condition and also by the observation that in the scalar-tensor picture of $f(R)$ gravity, the scalar curvature plays the role of the scalar field, and the continuity of $R$ is merely a statement of the fact that there is no membrane paradigm for a scalar field. 

We can also study the thermodynamics of the fluid membrane. We turn to that next.

\section{Thermodynamics of the membrane}
In order to study the thermodynamics of the membrane, we will assume that the spacetime is stationary with a timelike Killing vector which is null on the horizon. In the semiclassical limit, where the dominant contribution comes from classical field configurations, the partition function is
\bea
Z \approx \exp{\left(-(I^{\rm out}_E + I^{\infty}_E + I^{\rm surf}_E)\right)} = \exp{(-\beta F)} \; ,
\eea
where $I_E$ is the Euclidean action for the Euclideanized solution. Here $F$ is the free energy and $\beta$ is the periodicity of Euclidean time which is initially a free parameter but will ultimately be set to the inverse of Hawking temperature. The boundary term at infinity, $I_E^{\infty}$, is the appropriate generalization of the Gibbons-Hawking term, and is assumed to include any counter terms necessary to render the expression finite (such as a subtraction of the corresponding action for Minkowski space).
Now, in any stationary spacetime, the last term in (\ref{stresstensor}) vanishes in the null limit. The variation of the membrane action therefore reduces to
\bea
\delta I^{\rm surf}_{\Sigma} = -\frac{1}{8\pi G}\int d^{D-1}x \, \sqrt{-h} f'(R)  \lf K h^{ab} - K^{ab}\rt \delta h_{ab} \; .
\eea
We would like to integrate this variation to obtain an action for the membrane. 
If we set the variation of the Ricci scalar to zero on the horizon, we can easily integrate the membrane action \cite{membrane}. The result is
\bea
I^{\rm{surf}} = -\frac{1}{8\pi G} \int d^{D-1}x \, \sqrt{-h} f'(R)  K \; .
\eea
The condition $\delta R = 0$ on the horizon amounts to setting the variation of the scalar degree of freedom to zero in the scalar-tensor picture. In fact, the same condition is used on the external boundary to obtain the bulk equation of motion using a generalization of the Gibbons-Hawking surface term \cite{hill}. To evaluate the Euclidean action for the membrane, note that (\ref{nulllimit}) gives $K = \alpha^{-1} \kappa$ in a stationary spacetime, while Euclidean time, $t_E$, runs from $0$ to $\beta = 2 \pi/\kappa$. Hence
\bea
I_{E}^{\rm{surf}} =   \frac{1}{4 G} \int f'(R) \sqrt{\gamma}d^{D-2} x \; ,
\eea
where the sign change comes from the Wick rotation $dt = - i \, d t_E$. 
We can therefore calculate the entropy of the membrane:
\bea
S^{\rm{surf}} = \beta^2 \frac{\partial F}{\partial \beta} = -  \frac{1}{4 G} \int f'(R) \sqrt{\gamma}d^{D-2} x \; .
\eea
The membrane entropy is \textit{exactly} equal and opposite to the Wald entropy for $f(R)$ gravity \cite{Wald, Iyer}. If the entropy of the external universe is the same as the Wald entropy, then the entropy of the total system ``Membrane + External Universe" \textit{vanishes}. 
In the membrane approach, this suggests the following interpretation.
For an external observer, there is no black hole --- only a membrane. The entropy of the external world is then simply the total entropy of everything outside, which is equal and opposite to the entropy of the membrane. This number decreases as matter leaves the external system to fall through and
be dissipated by the membrane. When all matter has fallen through the membrane, the outside
is in a single state --- vacuum --- and has zero entropy.\\

To demonstrate this explicitly, consider the standard Schwarzschild spacetime of mass $M$ in four dimensions. This is still a solution of the vacuum $f(R)$ equations of motion. The stretched horizon can be taken to be simply a surface of constant Schwarzschild coordinate $r$. 
The bulk action vanishes on-shell and we find that
\bea
I_{E}^{\infty} = 4 \pi  G M^2 \; .
\eea
Hence, the entropy of the external universe is
\bea
S_{\textrm{ext}} =  4 \pi  G M^2 \; .
\eea
Now, for any polynomial $f(R)$ theory of the form $f(R) = R + ...$, we have $f'(R) = 1$ for this solution. In that case the entropy of the membrane is just
\bea
S_{\textrm{surf}} = - 4 \pi G M^2 \; .
\eea
The membrane entropy precisely cancels compensate the external entropy. As befits the generalized entropy, in a spacetime with no matter and no horizons, the total entropy is zero. That the membrane action reproduces (albeit with a properly interpreted minus sign) the black hole entropy is one of the advantages of the action formulation and one of the pleasing aspects of the membrane paradigm; it seems more satisfying that the horizon entropy can be traced to a term in the action that actually lives at the horizon, rather than at infinity.

\section{Summary}
We have extended the membrane paradigm for black hole horizons to general $f(R)$ gravity theories.
We have found that the membrane generically behaves like a non-Newtonian fluid with curvature-dependent transport coefficients; the dynamical equations of the membrane are identical to the corresponding equations in fluid dynamics adapted to a fluid with inhomogeneous and velocity-gradient-dependent viscosity coefficients.
We have also calculated the entropy of the membrane: it agrees with the suitable Wald entropy provided we set the variation of the Ricci scalar to zero on the horizon. 
Our calculations indicate that a membrane paradigm viewpoint may exist for general higher-derivative theories of gravity, but that there are subtleties, largely because there are additional physical degrees of freedom. It would be especially interesting to study the fluid properties of black hole horizons in Lovelock gravity, which has the same number of degrees of freedom as Einstein gravity.\\



\noindent
{\bf Acknowledgments}

\noindent
SS is supported by NSF grants PHY-0601800 and PHY-0903572. SS also thanks T. Jacobson for discussions.
\section{Appendix: $F_1$ term in (\ref{variationaction_g})}
\label{app:total-derivative}
We first note that any variations in the metric that are merely gauge transformations can be set to zero. Using a vector $v^a$ where $v^a$ vanishes on the stretched horizon, we can gauge away the variations in the normal direction so that  $\delta g_{ab} \to \delta h_{ab}$. Next we notice that for any vector $v^a \in \Sigma$, we have $\nabla_a v^a = v^{a}_{|a}$ where $_{|a}$ is the covariant derivative with respect to $h_{ab}$; integration over any divergence term like $v^{a}_{|a}$ over the stretched horizon gives zero. We also use relations like $h_{ab} n^b = 0$ and $a^d = n^e \nabla_e n^d = 0$. Then the integral of the $F_1$ term is
\begin{eqnarray}
\lefteqn{\int d^3 x \rh h^{bc} \lf \del _a \lf  f'(R)n^a \d h_{bc} \rt -
\del _c \lf f'(R) n^a \d h_{ab} \rt \rt } & & \nonumber \\
& = & \int d^3 x \rh \lf \del _a \lf h^{bc} f'(R) n^a \d h_{bc} \rt
+ \lf n^c a^b + n^b a^c \rt f'(R) \d h_{bc}
{}- \lf h^{bc} n^a f'(R) \d h_{ab} \rt _{|c} \right. \nonumber \\
& & \indent \indent \indent \left. - f'(R) h^{bc} n^a \d h_{ab} a_c
{}- K f'(R) n^b n^a \d h_{ab} - a^b n^a f'(R) \d h_{ab} \rt \nonumber \\
& = & \int d^3 x \rh \lf \del _a \lf h^{bc}  f'(R) n^a \d h_{bc} \rt
- K  f'(R) n^b n^a \d h_{ab} \rt \nonumber \\
& = & \int d^3 x \rh \lf \del _a \lf h^{bc} f'(R)
n^a \d h_{bc} \rt
- K f'(R)  \lf \d \lf n^b n^a h_{ab} \rt -
n^a h_{ab} \d n^b - n^b h_{ab} \d n^a \rt \rt \nonumber \\
& = & \int d^3 x \rh \del _a \lf h^{bc} f'(R) n^a \d h_{bc} \rt
\nonumber \\
\end{eqnarray}
Now let us take an auxiliary vector $k^a$ such that $n^a = U^a - \alpha k^a$. When $\alpha \to 0$, we have $ n^a \to U^a \to \alpha^{-1} l^a$ on the true horizon. Then the term $F_1$ ultimately becomes
\bea
F_1 &=& \int d^3 x \rh \del _a \lf h^{bc} f'(R) U^a \d h_{bc} \rt  - \alpha  \int d^3 x \rh \del _a \lf h^{bc} f'(R) k^a \d h_{bc} \rt  \; .
\eea
The second term does not contribute in the null limit and the integrand of the first piece is of the form $\nabla_a v^a = v^{a}_{|a}$ where $v^a \in \Sigma$. Hence, this term also does not contribute anything. This completes our proof that $F_1$ term in (\ref{variationaction_g}) vanishes in the null limit.





\newcommand{\eprint}[1]{\href{http://arxiv.org/abs/#1}{#1}}


\end{document}